\documentstyle[11pt,newpasp,epsf]{article}

\begin{document}

\title{Does Betelgeuse have a Magnetic Field?}

\author{S.B.F. Dorch\altaffilmark{1}} \affil{The Niels Bohr Institute for Astronomy, Physics and 
Geophysics, Juliane Maries Vej 30, DK-2100 Copenhagen {\O}} 
\author{B. Freytag} \affil{Department for Astronomy and Space Physics at Uppsala
University, Box 515, SE-75120 Uppsala, Sweden}

\altaffiltext{1}{Previously at the Institute for Solar Physics of 
the Royal Swedish Academy of Sciences, where the presented 
work was initiated.} 

\setcounter{footnote}{3}

\begin{abstract}
Betelgeuse is an example of a nearby cool 
super-giant that displays temporal brightness 
fluctuations and irregular surface structures. Recent numerical 
simulations by Freytag and collaborators of the outer convective
envelope comprising most of the entire star 
under realistic physical assumptions, have shown that the fluctuations 
in the star's apparent luminosity may be caused by giant cell 
convection, very dissimilar to solar convection. These detailed 
simulations bring forth the possibility of addressing another 
question regarding the nature of Betelgeuse and super-giants 
in general; namely whether these stars may harbor magnetic 
activity, which may contribute to their variability. 
Taking the detailed numerical simulations of the star at face 
value, we have applied a kinematic dynamo analysis to study 
whether or not the flow field of this super-giant may be able to 
amplify a weak seed magnetic field. We find that the giant cell 
convection does indeed allow a positive exponential growth rate of 
magnetic energy. The possible Betelgeusian dynamo can be 
characterized as belonging to the class of so-called ``local 
small-scale dynamos'' another often mentioned example of which is the 
dynamo action in the solar photosphere that may be 
responsible for the formation of small-scale flux tubes (magnetic 
bright points). However, in the case of Betelgeuse this 
designation is less meaningful since the generated magnetic field 
is both {\em global} and {\em large-scale}. 
\end{abstract}


\index{*Betelgeuse|alpha Ori}

\section{Introduction}

While the cool super-giant star Betelgeuse ($\alpha$ Orionis) is 
among the stars with the largest apparent diameters---corresponding 
to a radius in the range 600--800 ${\rm R}_{\odot}$---fundamental 
stellar parameters for this red M1--2 Ia--Iab star are by no means 
well-defined. 
Recently Freytag and collaborators (e.g.\ Freytag, 
Steffen, \& Dorch 2002) performed detailed numerical 
three-dimensional radiation-hydrodynamic simulations of the 
outer convective envelope and atmosphere of the
star under realistic physical assumptions. They try to determine if 
its observed brightness variations may be understood by convective 
motions within the star's atmosphere: the resulting models are 
largely successful in explaining the observations as a consequence 
of giant-cell convection on the stellar surface, very dissimilar 
to solar convection. These detailed simulations bring forth the 
possibility of solving another question regarding the nature of 
Betelgeuse, and of super-giants in general; namely whether these 
stars may harbor magnetic activity, which in turn may also 
contribute to their variability. A possible astrophysical dynamo 
in Betelgeuse would most likely be very different from those 
thought to operate in solar type stars, both due to its slow 
rotation, and to the fact that only a few convection cells are 
present at its surface at any one time. 

\section{Dynamo model}

The basic ansatz of the approach in this paper is that the input 
{\em prescribed} flow-field is taken at face value; i.e.\ that the 
velocity {\em ceteris paribus} 
represents the true quantity in the real
Betelgeuse. 

We solve the induction equation for the prescribed velocity field, 
i.e.\ within the kinematic MHD approximation, which is valid when 
the magnetic field is weak: 
\begin{equation} 
 \frac{\partial {\bf B}}{\partial t} = 
 \nabla \times ({\bf u} \times {\bf B}) + \eta \nabla^2 {\bf B}, \label{eq-1}
\end{equation}
where ${\bf B}$ is the magnetic field (flux density), ${\bf u}$ is 
the prescribed velocity field, and $\eta$ is the magnetic 
diffusivity (the resistivity). 

When ${\bf B}$ becomes comparable to the equipartition value of
the convective flows B$_{\rm eq} = {\rm u} \sqrt{\mu \rho}$, non-linear 
effects becomes important through the back-reaction of the Lorentz 
force on the flow. We are not able to model this non-linear 
behavior exactly within the present approach---the flow 
field is taken from a long since done calculation---instead we 
attempt an ad hoc strategy: as a first step towards including 
non-linearity, in a few cases we replace the flow ${\bf u}$ in Eq.\ 
(\ref{eq-1}) by a flow ${\bf u}_{\rm q}$ quenched by the magnetic 
field: 
\begin{equation} 
{\bf u}_{\rm q} = {\bf u} \exp -\alpha (e_{\rm m}/e_{\rm k})^2, 
\label{eq-2} 
\end{equation}
where $\alpha$ is a constant, $e_{\rm m}$ the magnetic 
energy density, and $e_{\rm k}$ is the average kinetic energy 
density, taken to be constant in both space and time. 

This quenching thus reduces the velocity amplitude at the 
locations where the magnetic energy becomes comparable to the 
kinetic energy of the fluid flow. Hence it reduces the growth rate 
of the magnetic field in these regions, causing the total magnetic
energy E$_{\rm m} = \int_{\rm V} {\rm e}_{\rm m} {\rm dV}$ to saturate. We 
have chosen $\alpha = 1.75$, which ensures the flow to be 
effectively unquenched until field strengths of $\approx 0.1~ 
{\rm B}_{\rm eq}$, while the flow amplitude 
completely vanishes at field strengths of $\approx 1.2~ {\rm B}_{\rm eq}$. 
Note that the constant $e_{\rm k}$ merely 
introduces a scaling factor in the solution to ${\bf B}$, and that 
its actual value is irrelevant to the present formulation of the 
problem. In future work the simple quenching Eq.\ (\ref{eq-2}) 
will be replaced with an expression taking into account the 
geometries of the field and flow.  

\subsection{Numerical method}

The employed numerical scheme for the kinematic dynamo simulations
is based on the staggered grid 
method by Galsgaard, Nordlund and others (e.g.\ Galsgaard \& 
Nordlund 1997): it uses 6th order staggering operators, 5th order 
centering routines, and a 3rd order Hyman predictor-corrector 
time-stepping. The same code has been used in the context of 
dynamo action previously, to study both kinematic dynamo action 
(Dorch 2000), as well as non-linear turbulent dynamos (Archontis, 
Dorch, \& Nordlund 2002), and recently to study kinematic dynamo 
action by convective flows in M-type dwarf stars (Dorch \& Ludwig 
2002). 

We implement a perpendicular magnetic field boundary condition by introducing
symmetry conditions on the magnetic field across ghost 
zones at the boundaries in all three dimensions. One could argue that ideally
radial or potential field boundary condition are more physical, and it is
indeed the plan to use such realistic boundary conditions in future work.
However, we have tested different simple boundaries and the results in 
terms of integral quantities such as e.g.\ magnetic energy seem quite 
robust.

Likewise, we have employed different types of initial conditions for the 
magnetic field and find that after a short transient, the behavior
of the system does not depend on the particular choice of topology, 
as it will become
apparent from the subsequent discussion of our results. The simulations are
initiated with either a unidirectional weak field, or a periodic weak field
with a large number of null-points (an ABC-like topology, see e.g.\
Dorch 2000).

\subsection{Flow field input data}

The input flow field results from the star-in-a-box models of co-author
Freytag and collaborators (see e.g.\ Freytag, Steffen, \& Dorch 2002).
The particular dataset used here is from a rather coarse model with
$127^3$ grid points; the model identification designation is 
{\tt dst33gm06n03}. 

The giant convection cells are so large that only a few cells are present at 
the surface at the same time; hence the situation is very different from
that of dwarf stars such as the Sun, where there can be thousands of cells
present at the surface. Betelgeuse is only slowly rotating and the star-in-a-box 
model does not include rotational forces: it follows that the flows are not
very helical. In fact the relative kinetic helicity $\langle {\bf \omega} \cdot 
{\bf u} \rangle / (\omega_{\rm rms} {\rm u}_{\rm rms})$ is only on the order of 
0.02 (where $\omega = \nabla \times {\bf u}$ is the vorticity).

The flow field data comprises 120 snapshots covering
7.5 years of giant cell convection. This is actually not the full duration of 
that particular simulation, but we had to limit the amount of input data, due 
to lack of available disk space. The time step resulting from the Hyman 
predictor-corrector scheme used when solving Eq.\ (\ref{eq-1}) is typically 
25 times smaller than the interval between the flow field snapshots; 
interpolation at each time step is hence necessary and sufficiently smooth
behavior is achieved using a simple first order interpolation routine.

To be able to study longer time sequences than the 7.5 year
extent of the 
input data, we cyclically re-use the input flow, effectively introducing
a 7.5 year periodicity in the flow; while such a periodicity arguably is
not observed, nor likely in Betelgeuse, it allows us to study long term
effects related to diffusion on global scales.

\begin{figure}[!htb]
\plotone{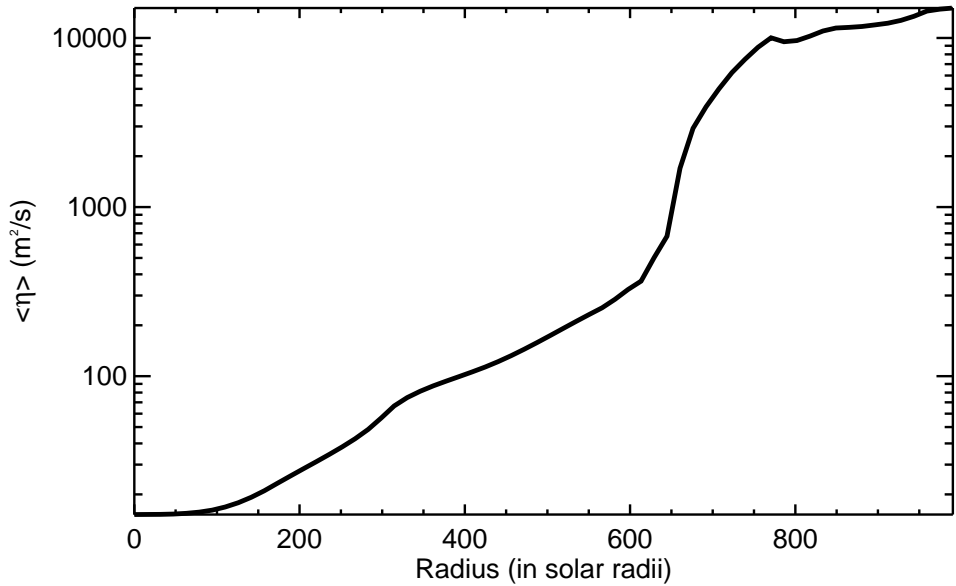} 
\caption{Diffusivity: average radial magnetic 
diffusivity $\eta$ (m$^2$/s) in the model of Betelgeuse as a 
function of radius in solar radii R$_{\odot}$.} \label{fig-1} 
\end{figure}

\subsection{Diffusion and magnetic Reynolds number}

Dynamo action by flows are often studied in the limit of 
increasingly large magnetic Reynolds numbers Re$_{\rm m} = \ell 
{\rm U}/\eta$, where $\ell$ and U are characteristic length and 
velocity scales. Most astrophysical systems are highly conducting 
(yielding small magnetic diffusivities/resistivities $\eta$), and 
their dimensions are huge; consequently values of Re$_{\rm m}$ are 
huge too. Seemingly odd exceptions are e.g.\ cool M-type dwarf stars 
(see Dorch \& Ludwig 2002) that have atmospheres a hundred times 
more neutral than the Sun. Betelgeuse is not an exception however; 
most parts of the star is better conducting than the solar surface 
layers, which has a magnetic diffusivity of the order of $\eta 
\approx 10^4$ m$^2$/s. 

Figure \ref{fig-1} shows the average Spitzer's resistivity as a 
function of radius in the model of Betelgeuse: Spitzer's formula 
(e.g.\ Schrijver \& Zwaan 2000) assumes complete ionization and 
hence the precise values of $\eta$ are uncertain in the outer 
parts of the star, where the atmosphere borders on neutral. There 
is some uncertainty connected also with defining the most 
important length scale of the system, but taking $\ell$ to be 10\% 
of the radial distance R from the center (a typical scale of the 
giant cells), and U $= {\rm u}_{\rm RMS}$ along the radial direction 
yields Re$_{\rm m}=10^{10}$--$10^{12}$ in the interior part of the 
star where R $\le 700$ R$_{\odot}$. 

Our numerical approach invokes quenching of the magnetic 
diffusivity by the convergence of the flow-field across the 
magnetic field, so that the magnetic Reynolds number Re$_{\rm m}$ 
is large in the bulk of the flow (this diffusion quenching should 
not be confused with the quenching of the flow given by Eq.\ \ref{eq-2}). 
The primary 
advantage of this approach is that diffusion is confined to the 
small regions where it is in fact needed, in order to resolve the 
smallest magnetic structures on the numerical grid (where the 
magnetic field gradients are large): typically we set the minimum 
value of Re$_{\rm m}$ larger than a few hundred. 

\begin{figure}[!htb]
\plotone{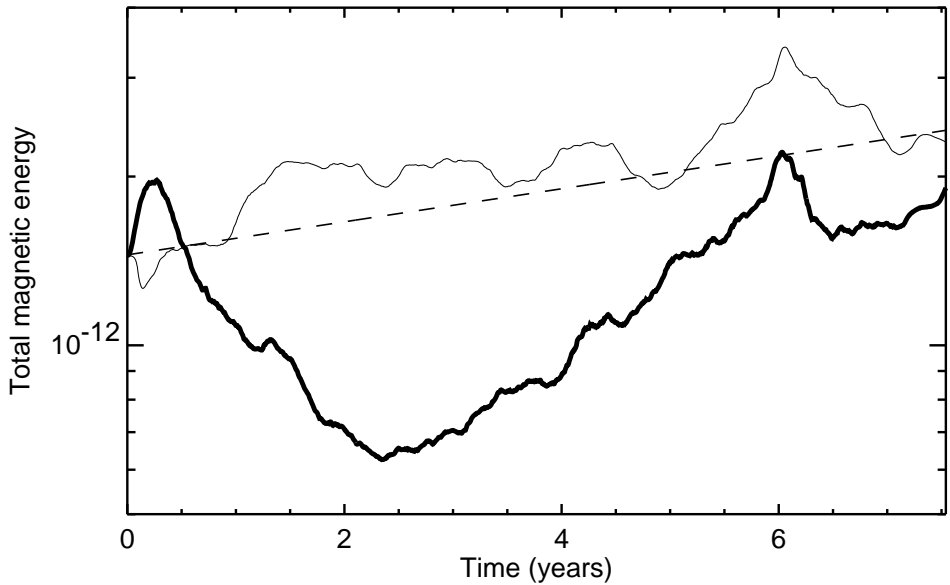} 
\caption{Initial growth: total magnetic energy E$_{\rm m}$ as 
function of time in years, for the simulation with Re$_{\rm m} \ge 
800$ (thick solid curve), and quenching switched off. 
Shown is the evolution during the time interval covered
by the flow field data, i.e.\ 7.5 years. Also shown is the average
energy over 9 subsequent periods of 7.5 years (thin curve). The
dashed line indicates growth with an e-folding time of 14.75 years.} \label{fig-0} 
\end{figure}

\section{Results}

There seems to be some disagreement as to what one should require 
for a system to be an astrophysical dynamo. Several ingredients can be 
considered to be necessary in order for a system to be a ``true''
dynamo; we believe that the following four should suffice. 
\begin{enumerate}
 \item The flows must stretch, twist and fold the magnetic field lines.
 \item Reconnection must take place to render the above processes irreversible 
 (i.e.\ diffusion is needed locally).
 \item The weak magnetic field must be circulated to the locations where 
 flow can do (Lorentz) work upon it.
 \item The total volume magnetic energy E$_{\rm m}$ must increase 
 (if a kinematic dynamo).
\end{enumerate} 
The conjecture that these necessary ingredients are sufficient is
based largely on our experience from idealized kinematic and 
non-linear dynamo models (Archontis \& Dorch 1999; Dorch 2000; 
Archontis, Dorch, \& Nordlund 2002). 

\begin{figure}[!htb]
\plotone{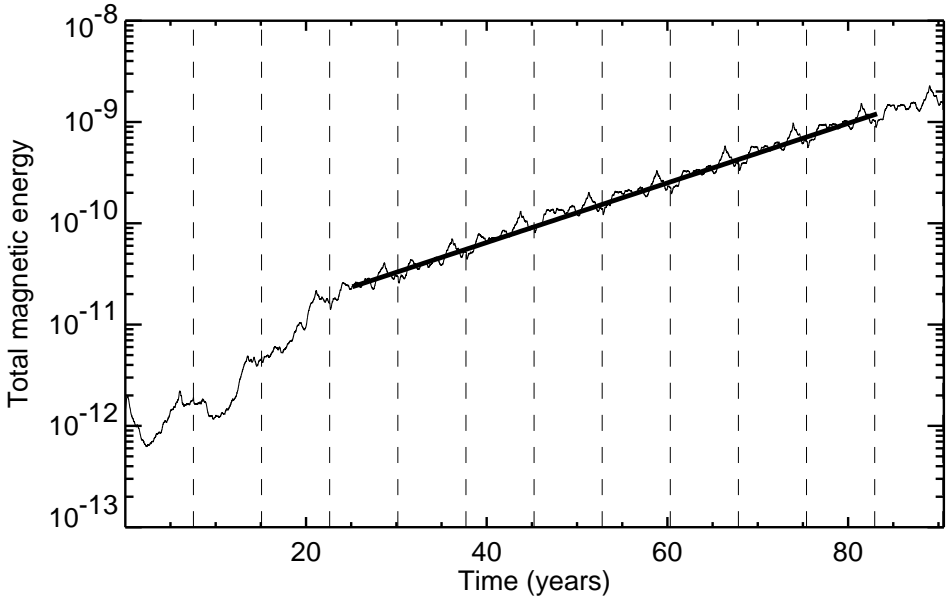} 
\caption{Kinematic dynamo: total magnetic energy E$_{\rm m}$ as 
function of time in years, for the simulation with Re$_{\rm m} \ge 
800$, quenching switched off, and cyclic re-use of 
the flow field (thin curve).
Also shown is a line (thick) indicating 
exponential growth corresponding to 14.7 years, and vertical dashed
lines indicating the periodicity of the flow field.} \label{fig-2} 
\end{figure}

The present contribution presents however only a preliminary study 
of a possible Betelgeusian dynamo, and we shall be dealing mainly 
with the last of these four ingredients: namely the question of 
exponential growth of E$_{\rm m}$. Identification of the mode of operation
of the dynamo will be presented elsewhere.

\begin{figure}[!htb]
\plotone{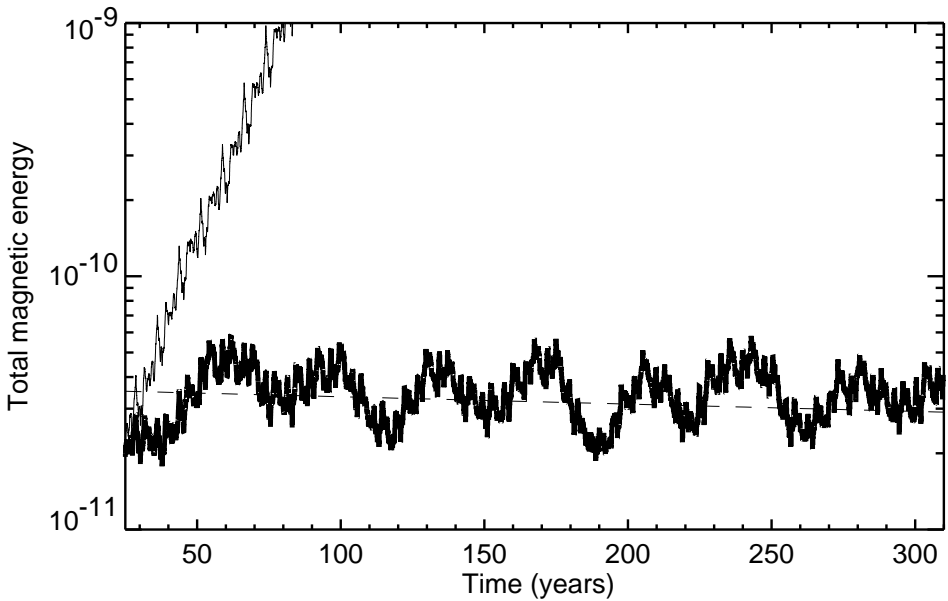}
\caption{Quenched dynamo: total magnetic energy E$_{\rm m}$ as 
function of time in years, for the simulation with Re$_{\rm m} \ge 
800$ and quenching of the flow field (thick curve).
Overplotted is a curve corresponding to an exponential decay with
an e-folding time of 1500 years (dashed curve).
For comparison the corresponding purely kinematic results are also shown
(thin curve).} \label{fig-3} 
\end{figure}

In general we obtain dynamo action when the specified minimum value of 
Re$_{\rm m}$ is larger than approximately 500: at lower values of 
Re$_{\rm m}$ the total magnetic energy decays throughout. 
In the following we
report on results for cases with $min({\rm Re}_{\rm m}) = 800$.
In that case Re$_{\rm m}$ is much larger (on the order of $10^{10}$) 
in the bulk of the flow. 

Figure \ref{fig-0} shows the evolution of the total volume magnetic energy
E$_{\rm m}$ during the 7.5 year time interval covered by the flow field 
data: initially there is a short increase in E$_{\rm m}$ during the first
half year, which is followed by a two year decline, and a return to
exponential growth in the reminder of the simulation. One could worry that
the particular choice of magnetic field initial condition would influence
the results, but as already hinted this is not the case, as is
evident in Figure \ref{fig-2}, which shows the evolution of E$_{\rm m}$
over 90 years. In this model, the flow field data was re-used 12 times
(patched together at the ends by interpolated). 
The first 4 recyclings happens to begin with different magnetic field 
configurations, and they can thus be thought of as
corresponding to simulations with different initial conditions: in all
cases there is an over-all exponential growth of the total magnetic energy. 
In the
last 9 cycles until the simulation was terminated, the growth of the 
magnetic field proceeds in a very regular (artificial, one might
add) manner, with a well defined average exponential growth rate corresponding
to a growth time of 14.7 years. 

In Figure \ref{fig-0} we have overplotted the average evolution of E$_{\rm m}$
in the last 9 cycles of the simulation: in these cycles the behavior of 
E$_{\rm m}$ is completely regular, and we speculate the it corresponds 
to some eigen-mode of the 
dynamo (determined by Eq.\ \ref{eq-2}), i.e.\ the field geometry at the 
beginning of each cycle is the one that belongs to the largest growth
rate, which at $min({\rm Re}_{\rm m}) = 800$ turns out to be 
$\sim 0.07$ yr$^{-1}$.

\begin{figure}[!htb]
\plottwo{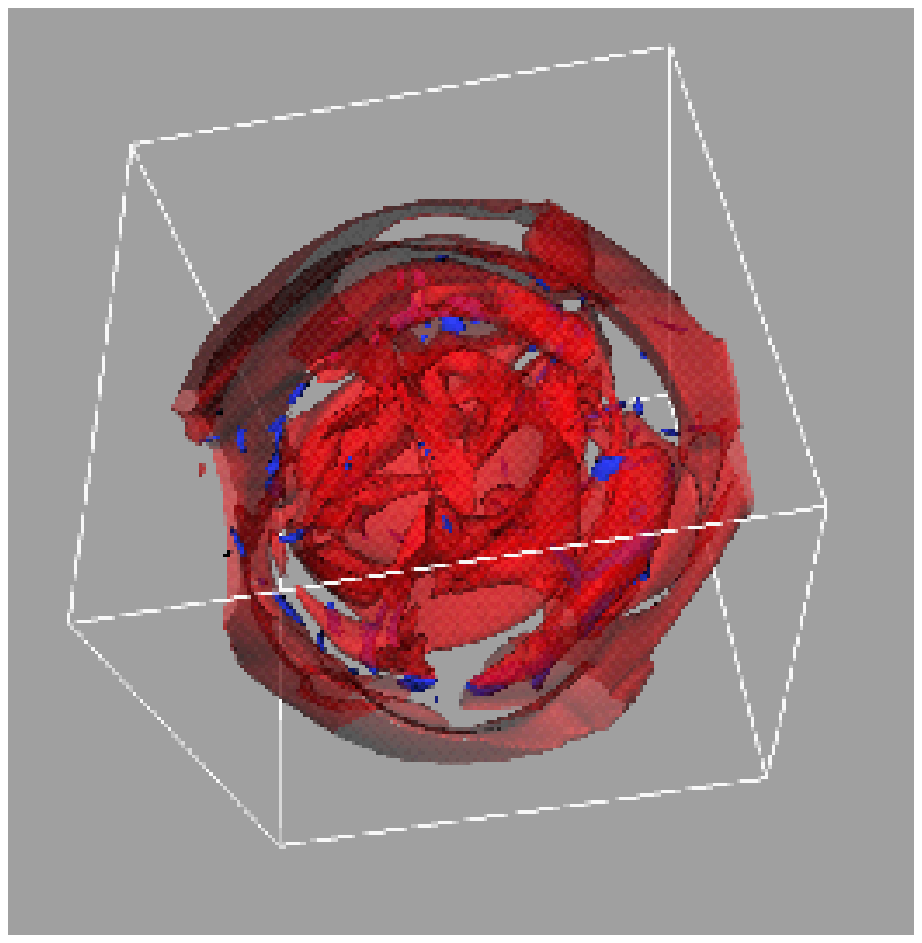}{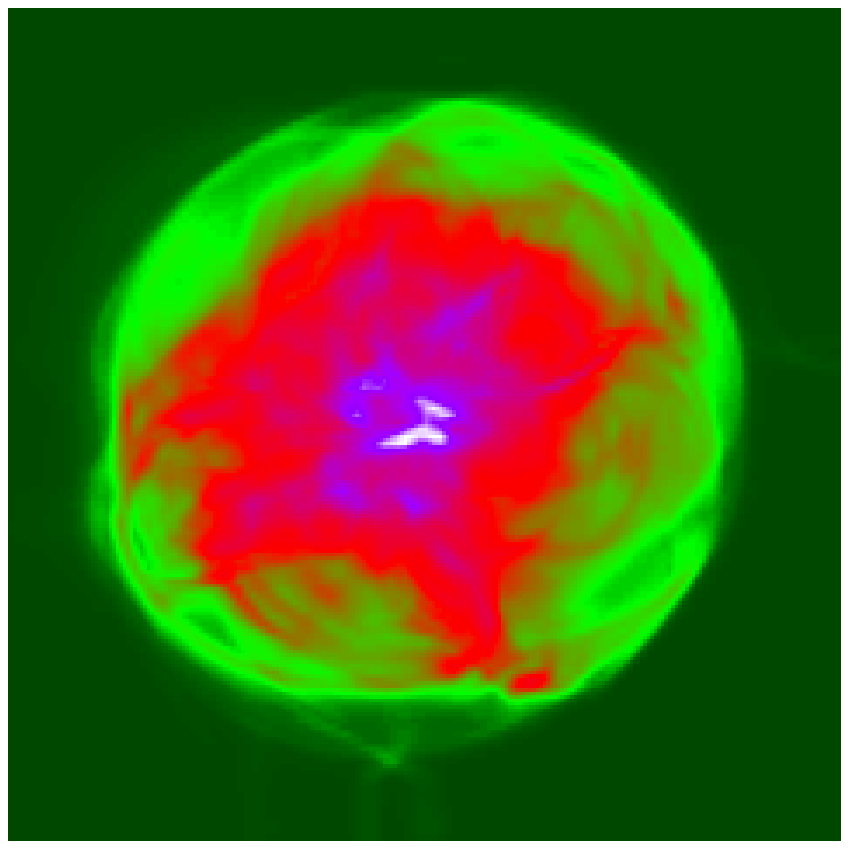} 
\caption{Magnetic structures. Left: a volume rendering of a snapshot 
of magnetic field 
strength (blue/dark isosurfaces) and velocity (red/transparent 
isosurfaces), from the simulation with $min({\rm Re}_{\rm m}) = 800$. The 
magnetic field strength isosurfaces are at a high value relative to the
maximum at that instant, and the velocity isosurfaces are at an
intermediate value. Right: an average image of the magnetic field obtained
by adding three images resulting from the average along the coordinate axis 
(white corresponds to maximum field strength, and (dark) green to minimum).} 
\label{fig-5} 
\end{figure}

No exponential growth can go on forever and eventually the magnetic energy
amplification must saturate. To model this non-linear effect, we ran a
simulation identical to the kinematic model yielding the result in 
Figure \ref{fig-2}, but including the flow quenching of Eq.\ (\ref{eq-2}). 
Figure \ref{fig-3} shows the evolution of a saturating dynamo, after an 
initial 25 year exponential growth (identical to that of the purely 
kinematic simulation): in that case the flow field cycles become less apparent
in the evolution of E$_{\rm m}$, and a additional long term variations
become visible---the dominant period seems to be about 35 years (more than
4.5 cycles). The simulation was terminated after 300 years in Betelgeusian
time.
One may worry that we have not run the simulation for long enough compared to some
dominate diffusion time scale. If diffusion takes place on a time scale related
to the largest scale of the system (see e.g.\ Brandenburg 2001), then the 
characteristic diffusion time would be $\tau_{\rm d} = {\rm R}^2 / 4 \pi^2
\eta \sim 2~ 10^{10}$ years, i.e.\ longer than the age of the Universe.
However, scaling this with our minimum Re$_{\rm m} = 800$ 
(to its actual value that is on the order of $10^{10}$)
yields a diffusion time of $\tau_{\rm d} \sim 1500$ years: that time scale does
not presently seem completely 
inconsistent with our results (see Figure \ref{fig-3}), but
the simulation covers only a fraction of $\tau_{\rm d}$.

The level of saturation of E$_{\rm m}$ is effectively set by the 
choice of the
constant kinetic energy density in Eq.~(\ref{eq-2}): thus the
model does not contribute any additional knowledge about which 
field strengths to expect in Betelgeuse.  
\begin{figure}[!htb]
\plottwo{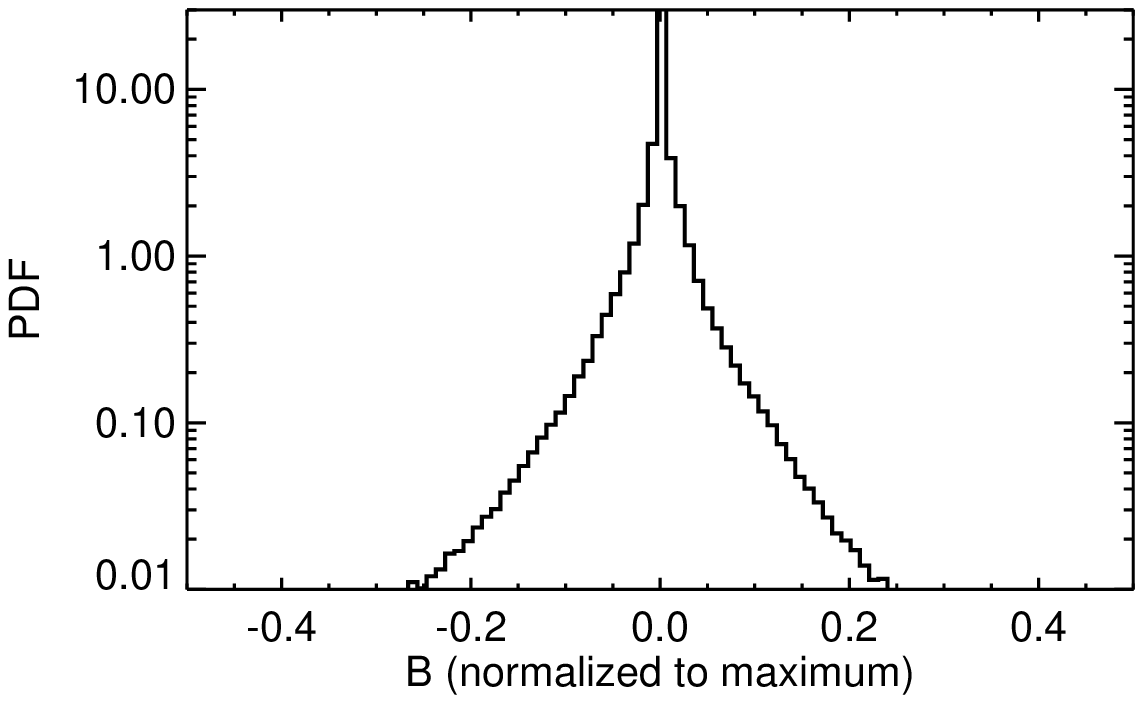}{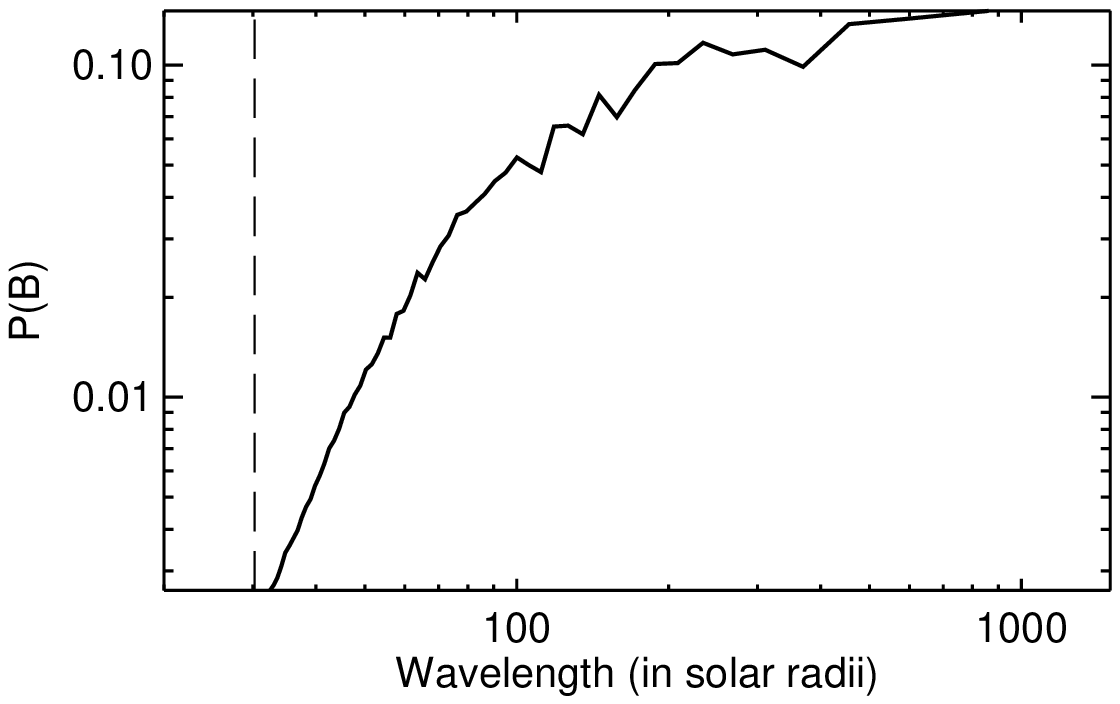} 
\caption{Intermittency. Left: the probability distribution function (PDF) for the 
magnetic field strength (normalized to its maximum) 
at an instant in the non-saturating kinematic model, 
corresponding to the snapshot in Fig.\ \ref{fig-5}. Right: power spectrum
of the magnetic field strength (full curve). The vertical dashed line indicates
the formal resolution limit (the Nyquist wavelength).} 
\label{fig-6} 
\end{figure}
But we can, however, study the geometry of the magnetic field that the 
dynamo generates. Figure \ref{fig-5} (left) shows a volume rendering of 
isosurfaces of magnetic field structures with a high field strength relative
to the maximum: the field becomes concentrated into elongated structures 
much thinner than
the scale of the giant convection cells, but perhaps due to the very dynamic
nature of the convective flows, no ``intergranular network'' is formed 
(to use solar terminology). The field is highly intermittent (see PDF in 
Figure \ref{fig-6}, left), i.e.\ only a small fraction of the volume carries the
strongest structures. The average magnetic field distribution is illustrated
in Figure \ref{fig-5} (right), where it is evident that the field is 
stratified and decreases in strength from the center of the star to a 
rather sharp ``magnetic surface'' at R $\sim 750$ R$_{\odot}$. We speculate
that the stratification results from the magnetic pumping effect 
working on the weakest part of the field (cf.\ Dorch \& Nordlund 2001).

The magnetic structures are well resolved, with maximum
power on the largest scales above 200 R$_{\odot}$ (Figure \ref{fig-6}, right). 
Additionally we observe a slight trend in the topology of the
field; fields near (but still below) the surface of the star are predominantly
horizontally aligned, while those in deeper layers are radial.

\section{Concluding remarks}

Based on the results presented here, we may not say conclusively if 
Betelgeuse {\em does} have a magnetic field, of course.
The results are tentative and should be used with caution.
But we may say that it seems that it {\em can} indeed 
have a presently unobserved magnetic field.
The dynamo of Betelgeuse may be characterized as belonging to the 
class called ``local small-scale dynamos'' another example of 
which is the proposed dynamo action in the solar photosphere that 
may possibly be responsible for the formation of small-scale flux 
tubes (cf.\ Cattaneo 1999, but also the discussion by Stein, ibid). 
However, in the case of Betelgeuse 
this designation is less meaningful since the generated magnetic 
field is both global and large-scale.

The future developments of this project will involve firstly using longer time
sequences of the input flow field, to avoid having to rely on recycling.
Secondly, simulations with higher numerical resolution is currently being 
carried out (see Freytag \& Finnsson, 2002, and Freytag \& Mizuno-Wiedner, 2002), 
which will allow larger runs with higher magnetic 
Reynolds numbers (i.e.\ smaller magnetic structures can form). 
Thirdly, it will be a priority to apply a more realistic quenching expression
to introduce 
the saturation (one that takes into account the relative inclinations
of the field and the flow, e.g.\ the cross-convergence). 
Lastly, more realistic boundary conditions, and a parameter study with increasing
Re$_{\rm m}$ will be performed.
The final goal is to be able
to identify the more of operation during the (non-linear) saturation phase
of the dynamo at high Re$_{\rm m}$ in order to predict the likely topology of the
magnetic field that one might observe at stars such as Betelgeuse.

\acknowledgments

SBFD thanks the LOC of IAU Symposium No.\ 210 for support. The
radiation-hydrodynamic simulations yielding the input flow field
used in this study was computed at the computing center of the
university in Kiel, at the Theoretical Astrophysics Center in
Copenhagen, and on the Sun cluster at {\AA}ngstr\"{o}m laboratory
in Uppsala. The dynamo calculations were performed at the 
Institute for Solar Physics of the Royal Swedish Academy of 
Sciences.

\end{document}